# Microscale selective laser sintering of Cu nanoparticles with a short-wavelength nanosecond laser

Youwen Liang, Bo Shen, Wan Shou*
Department of Mechanical Engineering, University of Arkansas, AR 72701, USA
*Corresponding author: wshou@uark.edu

**Abstract**

Microscale additive manufacturing of reflective copper is becoming increasingly important for micro-electronics and micro-computers, due to its excellent electrical and thermal conductivity. Yet, it remains challenging for state-of-the-art commercial metal 3D printers to achieve sub-100-micron manufacturing. Two aspects are sub-optimal using commercial laser powder bed fusion systems with infrared (IR) lasers (wavelength of 1060-1070 nm): (1) IR laser has a low absorption rate for Cu, which is energy-inefficient for manufacturing; (2) short wavelength lasers can potentially offer higher resolution processing due to the diffraction-limited processing. On the other hand, laser sintering or melting typically uses continuous wave (CW) lasers, which may reduce the manufacturing resolution due to a large heat-affected zone. Based on these facts, this study investigates the UV (wavelength of 355 nm) nanosecond (ns) laser sintering of Cu nanoparticles. Different laser processing parameters, as well as different nanoparticle packing densities, are studied. Our results show that a short-wavelength laser can reduce the required energy for sintering with decent morphology, and a densified nanoparticle powder bed favors continuous melting. We further show that sub-20 micron printing can be readily achieved with a UV ns laser. These findings provide new insights into short-wavelength laser–metal nanoparticle interactions, which may pave the way to achieve high-resolution micro/nano-scale additive manufacturing.

## Introduction

Micro- and nanoscale structures made from inorganic functional materials are increasingly vital in electronics, photonics, robotics, and computing devices. Realizing additive manufacturing (AM) of pure inorganic materials, such as metals, at the microscale could enable major technological breakthroughs. These include enhanced thermal management in electronics [1–4], compact system integration [5–9], novel MEMS architectures [9,10], and emerging applications such as terahertz (THz) communication [11–13], quantum computing [14–16], and nuclear energy [17–19].

Over the past two decades, AM has developed rapidly at the macro- and mesoscale using a range of materials, including polymers [20,21], metals [22,23], ceramics [24–26], and composites [27,28]. Yet, extending AM to the microscale (with features in the micron to tens-of-microns range) remains a critical frontier. Unlike top-down fabrication methods such as focused ion beam (FIB) milling [29] or lithography [30], micro-AM enables maskless, material-efficient 3D structuring. Various mechanisms have been developed to achieve micro- and sub-micron AM, including electrohydrodynamic printing [31,32], aerosol jet printing [33,34], direct ink writing (DIW) [35–37], femtosecond laser photoreduction [38–40], laser-induced forward transfer (LIFT) [41,42], and two-photon polymerization [43-45]. Among these, polymer-based micro-AM is the most mature, primarily using light-based approaches [46,47]; however, such systems are costly, and there is still no affordable, well-established AM solution for microscale metal fabrication [48–51]. Most commercial metal AM platforms, including selective laser sintering (SLS) and binder jetting, struggle to achieve sub-100 µm resolution. Yet, enabling features below this threshold could unlock transformative structural and functional applications [1–16].

Among various AM approaches, SLS or more broadly, laser powder bed fusion (LPBF) is promising for metals due to their scalability and versatility. However, its resolution is fundamentally limited by the diffraction limit [52]. According to Rayleigh's criteria [52-55], $d = \frac{1.22\lambda}{NA}$, where λ is the wavelength of light, and NA is the numerical aperture (NA) of the objective lens. Increasing NA using proper optical components or decreasing λ will lead to a smaller spot size or higher printing resolution. Current SLS/SLM systems typically use infrared (IR) lasers (1060–1070 nm), but substituting with ultraviolet (UV) lasers (355 nm) could theoretically improve resolution by a factor of three.

In addition to optical constraints, powder particle size and layer thickness impose further limits. Commercial LPBF systems generally use 20–50 µm powders with 20–100 µm layer thickness, yielding resolutions of 100–200 µm [56]. Recent efforts to reduce laser spot size (via high-NA optics), powder size, and layer thickness have achieved sub-100 µm features in metals [57-62]. Systematic studies on Cu nanoparticles using various lasers [49,63,64] have achieved sub-5 µm features, and flexible electronics applications have demonstrated 1–10 µm resolution with continuous-wave (CW) lasers at 514–532 nm [65-67]. Nevertheless, ablations were dominated by short-pulse laser processing [49,64]. Overall, smaller beam diameter, shorter wavelength, and finer powders can enhance printing resolution [50,68,69].

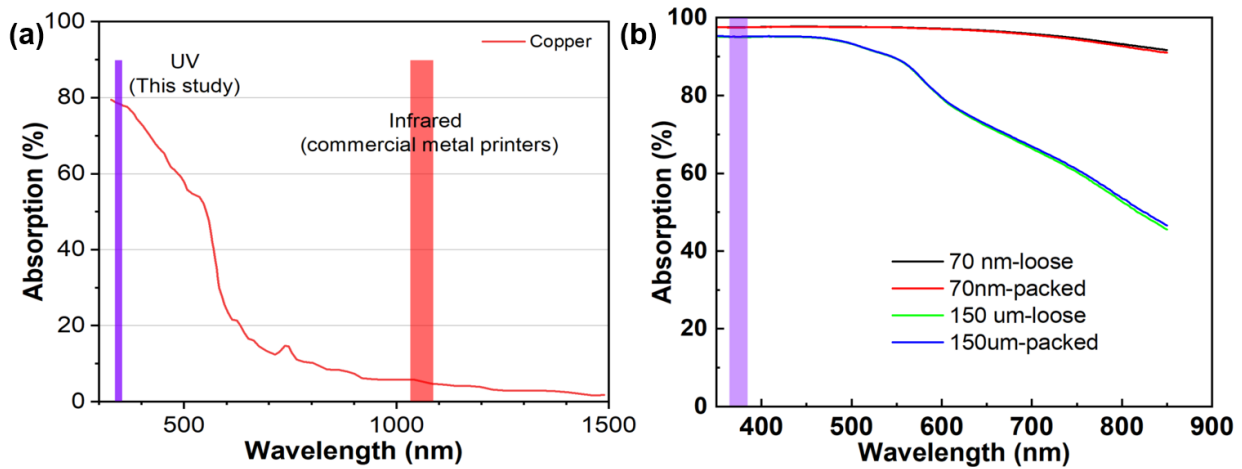

Fig. 1 (a) Absorption of solid copper at different wavelengths. Data extracted from [70]. (b) Comparison of the absorption of nanoparticles and microparticles.

Another key technical opportunity in microscale SLS lies in the spectral selectivity of laser–material interactions. Because lasers are monochromatic, their coupling with materials strongly depends on wavelength-specific absorption characteristics (as shown in Fig. 1). Therefore, selecting an optimal wavelength for a given material—considering both absorption and penetration depth—is critical for efficient energy utilization. Most commercial LPBF systems rely on IR lasers due to cost and legacy factors; however, IR wavelengths are poorly absorbed by many metals, including copper, silver, and gold [70]. In contrast, UV lasers can enhance absorption efficiency by more than an order of magnitude, potentially reducing energy consumption by factors of 15–30. Despite this advantage, the light–matter interactions at short wavelengths, particularly in the context of nanoparticle sintering, remain poorly understood.

In this paper, we investigate nanosecond (ns) UV laser selective sintering of copper nanoparticles at the microscale. This study aims to evaluate the feasibility of leveraging affordable ns laser systems to achieve high-resolution and energy-efficient metal additive manufacturing, while elucidating the underlying mechanisms that govern sintering behavior at this scale.

**Materials and Methods**

In this study, a nanosecond (ns) UV laser with a wavelength of 355 nm is used. The laser processing system (as shown in Fig. 2) is equipped with an objective lens of NA=0.65, which gives a laser spot size of about 5 µm. Copper (Cu) nanoparticles with a diameter of 70 nm are purchased from US Nano Research Nanomaterials, Inc. To realize a uniform nanoparticle film, nanoparticles are electro-sprayed on an aluminum foil with a powder coating gun. A proper pressure (20 psi) is used to condense the powder bed [71]. The absorption of nanoparticles at different wavelengths was measured with Fourier Transform Infrared Spectroscopy (FTIR), and compared with Cu microparticles, as indicated in Fig. 1b.

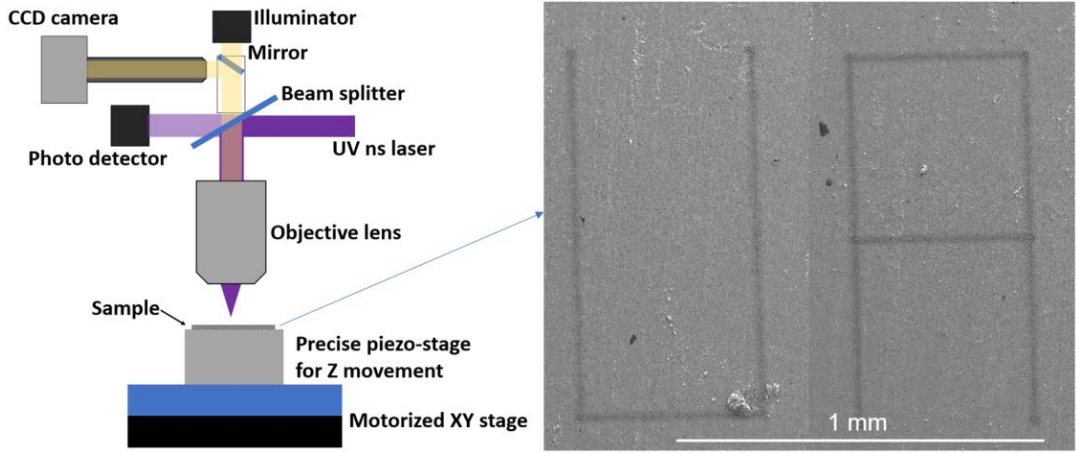

Fig. 2 Schematic drawing of the laser processing system (left) and a representative sintered "UA" pattern with microscale linewidth (right).

## Results and Discussions

### Pressing effect

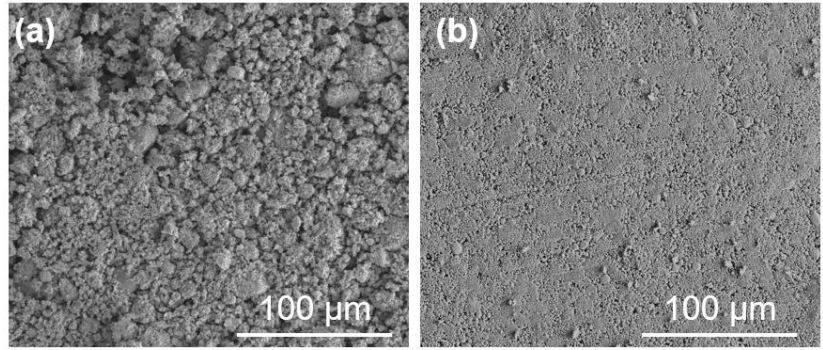

Fig. 3 Morphology of nanoparticle powder bed before and after pressing.

In this study, we first studied the consolidation effect through pressing before sintering. As shown in Fig. 3, the morphology of the nanoparticle film significantly changed from random and porous, to uniform and flat, which is similar to a previous study [71]. Subsequently, proper laser processing was conducted to compare the influence of pressing on the resulting morphology. As shown in Fig. 4, a general trend is observed, with the consolidation of the powder bed, the sintered line is flatter and more continuous. In contrast, due to the nonuniformity of the as-sprayed powder, the melted powders coalesce more randomly with their surrounding particles.

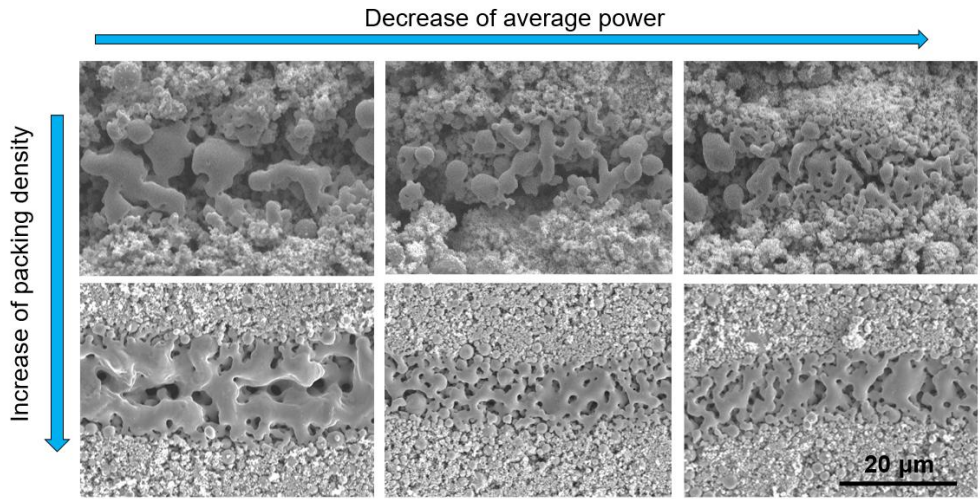

Fig. 4 The influence of packing density on sintered morphology (scanning speed 5 mm/s).

**Effect of laser processing parameters**

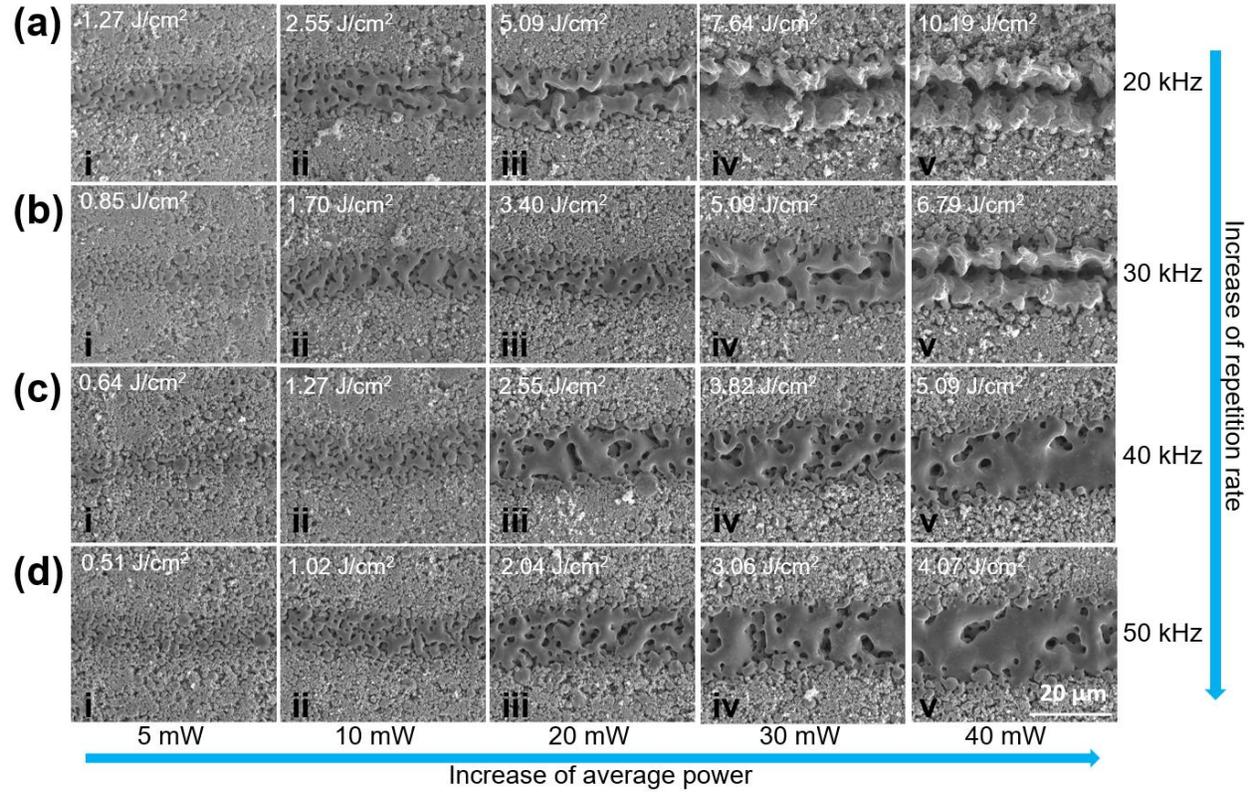

Fig. 5 UV ns laser sintering of Cu NP powder with various frequencies and fluences (scanning speed of 5 mm/s).

Table 1. Peak power (W) of different combinations of average power and repetition rate used in Fig. 5.

| Repetition rate (kHz) | Average power (mW) | | | | |
|---|---|---|---|---|---|
| | 5 | 10 | 20 | 30 | 40 |
| 20 | 22.32 | 44.64 | 89.29 | 133.93 | 178.57 |
| 30 | 12.17 | 24.33 | 48.66 | 72.99 | 97.32 |
| 40 | 7.72 | 15.43 | 30.86 | 46.30 | 61.73 |
| 50 | 5.19 | 10.38 | 20.76 | 31.14 | 41.52 |

Based on the above observation, we conducted a systematic study to understand the influence of laser processing parameters on the Cu nanoparticle sintering. As indicated in Fig. 1b, Cu nanoparticles have a high absorption of UV light; thus, a low average power is expected for the sintering. In our study, average power was tuned from 5 mW to 40 mW, corresponding to fluences ranging from 0.51 J/cm$^2$ to 10.19 J/cm$^2$ (as labelled in Fig. 5). The corresponding peak powers of various average power and repetition rate combinations are presented in Table 1. Figure 5 shows that, as laser power increases, low repetition rate irradiation (Fig. 5a), which corresponds to high peak power, leads to a transition from partial coalescence (Fig. 5a-i) to ablation (Fig. 5a-iii to v). In contrast, under high repetition rate conditions (Fig. 5d), which produce lower peak power, the surface morphology evolves from partial coalescence (Fig. 5d-i) to smooth, condensed melting (Fig. 5d-v). Generally, with the increase of repetition rate (from Fig. 5 a to d), the pulse width increases while the peak power decreases, which leads to a more gentle, uniform heating but a bigger heat-affected zone (HAZ). Correspondingly, the sintered results become flatter and more continuous (such as Fig. 5c-v and Fig. 5d-v). It was also noticed, with the same fluence (Fig. 5 a-iii, b-iv, and c-v), by tuning the repetition rate, the morphology can change from ablation to smooth sintering; however, the processed linewidth decreases with a low repetition rate (Fig. 5a-ii and 5c-iii). This is because with a low repetition rate, the pulse width is shorter, which limits heat conduction. Generally, our results are consistent with previous observations using green and blue lasers [70]; yet, our results demonstrate that ns pulsed laser sintering of pure Cu nanoparticles can be as promising as CW lasers [49,64]. Furthermore, by tuning the laser scanning speed and power with a fixed repetition rate of 40 kHz, we show that relatively smooth line with resolution of ~12 µm can be obtained (Fig. 6). Although the nanoparticles are consolidated to enhance the packing density, there are still some random holes in the sintered line (Fig. 6), which may require future study and optimization.

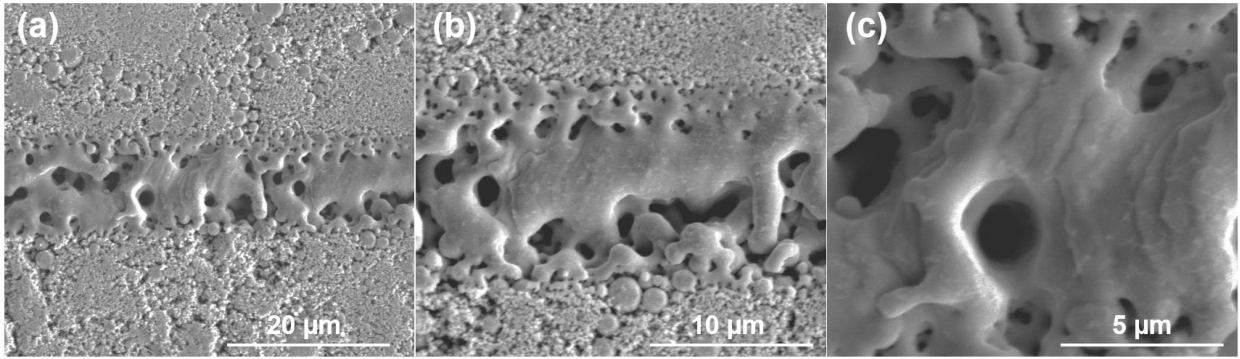

Fig. 6 Representative sub-20 µm sintering of Cu line (40 kHz).

In addition, we notice that in our processing, an extremely low average power, ≤40 mW (corresponding to a peak power of up to ~41.5 W), which is significantly lower than typical LPBF (hundreds of W) [12]. If only comparing the required average power of the UV ns laser to the power of widely used LPBF, the energy saving can be in 3 orders of magnitude. If we compare the peak power of the UV ns laser with the required power in LPBF, the energy saving is no less than 50%. No matter which metric is used, our results demonstrate the advantage of energy efficiency of using short wavelengths for metal AM. Here, the high-energy efficient manufacturing is mainly attributed to the high absorption percentage of Cu nanoparticles in the UV wavelength range.

## Focal plane influence

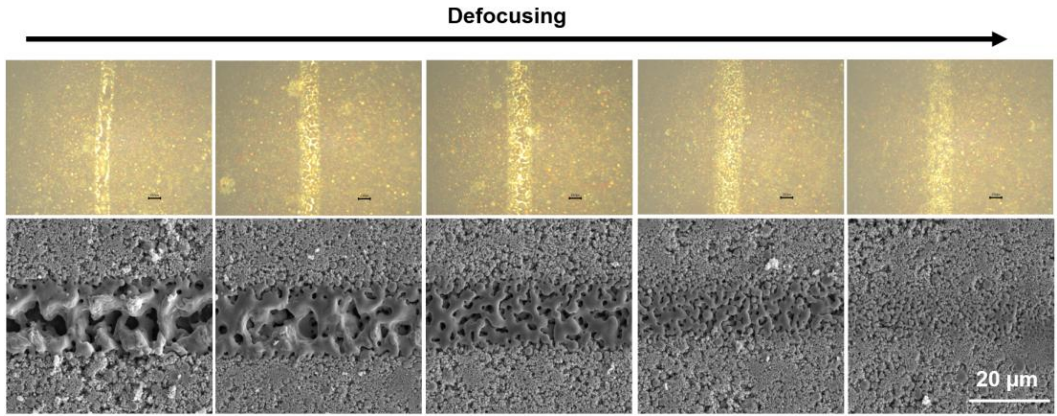

Fig. 7 The influence of the focal plane on the processed morphology. (Repetition rate 20 kHz, average power 5 mW, scanning speed 5 mm/s)

Lastly, we noticed that by slightly adjusting the focal plane, the ablation induced by ns laser processing due to high peak power, can be transitioned to smooth sintering (as shown in Fig. 7). Namely, by defocusing the laser spot (i.e., moving away from the focal plane), ablation can be eliminated. Yet, it should be noted that a defocused laser beam may not have sufficient power to melt the Cu nanoparticles (although with coalescence). Also, a defocused beam may reduce the printing resolution.

## Simulation study

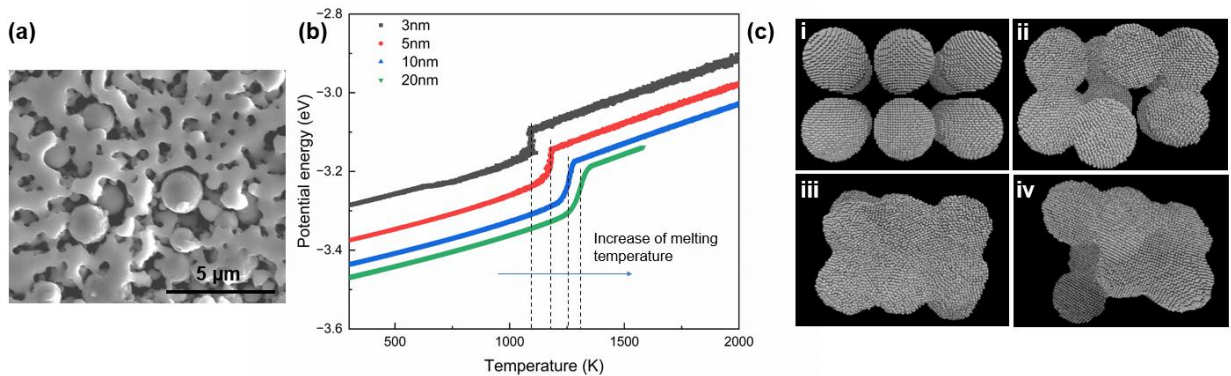

Fig. 8 (a) SEM observation of coalescence of nanoparticles with randomly distributed microparticles. (b) MD simulation of size effect on Cu nanoparticle melting; (c) coalescence of twelve 10 nm Cu nanoparticles below melting point at 1000 K: (i-iii) top view; (iv) size view.

In parallel, we conducted molecular dynamics (MD) simulations of melting with different Cu nanoparticle diameters to understand the size effect on sintering. Fig. 8a shows that the so-called 70 nm Cu nanoparticles have a wide range of sizes, and even with microscale particles, which is also observed by other researchers [63]. Such a particle diameter distribution will eventually affect the sintering morphology and printing quality. To gain a better understanding of how nanoparticle size affects melting or sintering, four different Cu nanoparticles are simulated for melting. As shown in Fig. 8b, the melting point increases from 1090 K to 1300 K when the nanoparticle diameter increases from 3 nm to 20 nm. This observation suggests that laser processing with Cu nanoparticles can use a much lower average power compared with using micro-powders. More interestingly, our MD simulations also suggest that coalescence can happen far below the melting temperature (Fig. 8c). This observation is consistent with previous experimental and simulation studies on gold, silver, and copper nanomaterials, which demonstrate spontaneous coalescence or welding at room temperature due to active atomic diffusion [72-74]. Such behavior is critical for minimizing the heat-affected zone and enabling high-resolution printing.

**Conclusions**

In summary, our systematic study of UV ns laser sintering of pure Cu nanoparticles demonstrates: (1) pulsed lasers can be promising for selective laser sintering using nanoparticles without ablation; (2) short wavelength lasers can significantly reduce energy consumption for metal AM; and (3) sub-20 µm printing of pure metal is feasible with nanoparticles and short wavelength lasers. We also show that proper condensation of the powder bed enhances the continuity of the sintered structures. Although this work focuses on Cu nanoparticles, the advantages of short-wavelength lasers—namely energy efficiency and high-resolution capability—are broadly applicable to other metal powders. Future work is needed to develop a more comprehensive understanding of the processing–morphology landscape, optimize printing resolution, and elucidate in greater detail the nanoparticle evolution during sintering.

**Acknowledgements**

This work is supported by the startup funding from the University of Arkansas. We thank Prof. Bo Zhao from the University of Houston for the assistance with UV-VIS spectrum measurement.